\documentclass{ieeeaccess}
\usepackage{cite}
\usepackage{amsmath,amssymb,amsfonts}
\usepackage{algorithmic}
\usepackage{graphicx}
\usepackage{textcomp}

\usepackage{booktabs}

\usepackage{bm}
\makeatletter
\AtBeginDocument{\DeclareMathVersion{bold}
\SetSymbolFont{operators}{bold}{T1}{times}{b}{n}
\SetSymbolFont{NewLetters}{bold}{T1}{times}{b}{it}
\SetMathAlphabet{\mathrm}{bold}{T1}{times}{b}{n}
\SetMathAlphabet{\mathit}{bold}{T1}{times}{b}{it}
\SetMathAlphabet{\mathbf}{bold}{T1}{times}{b}{n}
\SetMathAlphabet{\mathtt}{bold}{OT1}{pcr}{b}{n}
\SetSymbolFont{symbols}{bold}{OMS}{cmsy}{b}{n}
\renewcommand\boldmath{\@nomath\boldmath\mathversion{bold}}}
\makeatother

\def\BibTeX{{\rm B\kern-.05em{\sc i\kern-.025em b}\kern-.08em
    T\kern-.1667em\lower.7ex\hbox{E}\kern-.125emX}}

\begin{document}
\history{Received 24 November 2025, accepted 24 December 2025,date of publication 1 January 2026,\\
date of current version 6 January 2026}
\doi{10.1109/ACCESS.2025.3650352}

\title{Interactive Narrative Analytics: Bridging Computational Narrative Extraction and Human Sensemaking}

\author{\uppercase{Brian Keith}\authorrefmark{1}}

\address[1]{Universidad Católica del Norte, Antofagasta, Chile (e-mail: brian.keith@ucn.cl)}

\tfootnote{This research is funded by the ANID FONDECYT 11250039 Project. The author is also supported by Project 202311010033-VRIDT-UCN.}

\markboth
{Keith: Interactive Narrative Analytics: Bridging Computational Narrative Extraction and Human Sensemaking}
{Keith: Interactive Narrative Analytics: Bridging Computational Narrative Extraction and Human Sensemaking}

\corresp{Corresponding author: Brian Keith (e-mail: brian.keith@ucn.cl).}

\begin{abstract}
Information overload and misinformation create significant challenges in extracting meaningful narratives from large news collections. This paper defines the nascent field of \textbf{Interactive Narrative Analytics} (INA), which combines computational narrative extraction with interactive visual analytics to support sensemaking. INA approaches enable the interactive exploration of narrative structures through computational methods and visual interfaces that facilitate human interpretation. The field faces challenges in \textit{scalability}, \textit{interactivity}, \textit{knowledge integration}, and \textit{evaluation standardization}, yet offers promising opportunities across news analysis, intelligence, scientific literature exploration, and social media analysis. Through the combination of computational and human insight, INA addresses complex challenges in narrative sensemaking.
\end{abstract}


\titlepgskip=-21pt

\maketitle

\section{Introduction}
The digital landscape presents an unprecedented volume of information, with news and social media content growing exponentially \cite{gandomi2015beyond,roetzel2019information}. This information overload is compounded by the proliferation of misinformation and disinformation \cite{lazer2018science, vosoughi2018spread,zhou2020survey,apuke2024information}, creating a complex  ecosystem that challenges our ability to make sense of important events and their connections. Traditional information processing approaches have become inadequate \cite{eppler2004concept,cohen2011computational}, highlighting the need for more sophisticated computational and interactive tools to extract meaningful narratives from large text collections \cite{keith2023survey,dalkir2020navigating}.

This paper introduces and defines a new field of research that we call \textbf{Interactive Narrative Analytics} (INA), a multidisciplinary approach combining computational narrative extraction \cite{keith2023survey,ranade2022computational,santana2023survey} with interactive visual analytics \cite{andrews2010space,endert2011observation,endert2012semantic} to support sensemaking \cite{pirolli1995information,pirolli2005sensemaking}. By coining this term, we seek to establish a distinct research domain focused on addressing the unique challenges of narrative understanding in the digital age. Unlike traditional text analytics, which often focuses on statistical patterns or isolated entities, INA emphasizes temporal, causal, and relational aspects of information, capturing how events unfold and connect over time to form coherent stories \cite{keith2023survey}.

The need to establish INA as a field has become increasingly evident as analysts, journalists, researchers, and the public struggle with information complexity. The sheer volume of event data makes manual analysis virtually impossible \cite{cohen2011computational,lewis2013content,park2019does,soroya2021information}, necessitating scalable computational approaches. The dynamic nature of news events requires methods that can adapt to evolving narratives in real-time. The complexity of narratives demands sophisticated representation models capable of capturing intricate relationships. Furthermore, the subjective nature of narrative interpretation requires interactive tools that can incorporate human feedback and domain knowledge \cite{heer2012interactive,segel2010narrative,wenskovitch2020interactive, wenskovitch2021beyond}.

Current narrative extraction approaches employ techniques from natural language processing, machine learning, and graph theory \cite{shahaf2010connecting, keith2020maps, angus2023discourse}, but fall short in critical areas. Most existing methods lack scalability, struggling to process the massive volumes of data generated in today's information ecosystem. They typically operate as black boxes with limited transparency and interactivity, failing to leverage human intuition and domain expertise. Moreover, they rarely incorporate external knowledge, leading to incomplete or inaccurate narrative representations. Most importantly, they often lack effective evaluation metrics, making it difficult to compare different approaches \cite{keith2023survey, vivacqua2019explanations}.

This proposed field would bring together computational optimization, large language models (LLMs), and knowledge representation \cite{zhao2024leva, chang2024survey, yang2024give} with interactive visualization techniques \cite{han2023explainable, bian2021deepsi} to create an integrated approach to narrative understanding. While individual elements of this field exist in isolation across various research domains, we argue that establishing INA as a distinct discipline would foster the cross-disciplinary collaboration needed to address the complex challenges of narrative sensemaking.

In particular, this paper seeks to \textbf{define the scope of this nascent field}, examining its theoretical foundations, potential components, anticipated research challenges, promising applications, and future research directions. As a position paper proposing a new interdisciplinary field, we synthesize key concepts from multiple disciplines rather than providing a systematic literature review. By defining this new area of research, we aim to highlight the importance of integrating computational extraction with human sensemaking capabilities, providing a conceptual framework for future research in INA. For detailed surveys of the underlying technical areas, we refer interested readers to Keith et al. \cite{keith2023survey} on narrative extraction, Santana et al. \cite{santana2023survey} on narrative processing, and Keim et al. \cite{keim2008visual} on visual analytics foundations.

\section{Defining Interactive Narrative Analytics}
INA represents an emerging interdisciplinary field at the intersection of computational narrative extraction, visual analytics, and human-centered sensemaking. We define INA as the systematic approach to \emph{extracting, representing, and exploring event-based narrative structures} from large textual collections through a combination of computational methods that identify temporal, causal, and semantic relationships between events, interactive visual interfaces that support human exploration and refinement of these structures, and knowledge integration mechanisms that enhance narrative understanding with external domain knowledge.

In particular, INA focuses primarily on event-based narratives, particularly those found in news articles, social media, and similar domains where understanding temporal and causal relationships between events is crucial \cite{keith2023survey}. It emphasizes the interactive aspect of narrative analysis, acknowledging that narrative understanding requires human interpretation \cite{wenskovitch2020interactive}, and incorporates knowledge integration to enhance extracted narratives with external domain knowledge \cite{yan2023narrative}. Thus, unlike traditional text analytics approaches that focus on statistical patterns or isolated entities, INA is concerned with the holistic structure of narratives, including events, entities, temporal relationships, and causal connections. It extends beyond traditional narrative extraction by incorporating interactive visualization and human feedback into the extraction process itself, rather than treating these as separate downstream tasks \cite{keith2023iui}.

\subsection{Distinguishing INA from Related Fields}
To clarify INA's unique position in the research landscape, we must distinguish it from several related but distinct areas. Visual Analytics (VA) is defined as "the science of analytical reasoning facilitated by interactive visual interfaces" \cite{thomas2005illuminating}. VA emerged from intelligence needs following the 9/11 attacks \cite{thomas2005illuminating,keim2008visual} and provides the interactive visualization foundation upon which INA builds. However, while VA broadly addresses all data types and analytical tasks, INA specifically focuses on \emph{narrative structures} within textual data, emphasizing temporal event sequences and causal relationships rather than general data patterns. This distinction is crucial: VA might visualize any patterns in document collections, but INA specifically seeks to extract and represent the stories those documents tell.

Intelligence and security analysis applications, which historically drove VA development \cite{keim2008visual}, employ visual analytics primarily for threat detection, anomaly identification, and real-time monitoring. While INA can support intelligence applications, it extends far beyond security contexts to support narrative understanding across diverse domains including scientific literature, news media, and legal documents. The key difference lies in the analytical focus: intelligence analysis prioritizes identifying threats and anomalies, while INA emphasizes understanding narrative coherence, evolution, and the relationships between different storylines.

Event-based text mining represents another related area that shares INA's interest in events but differs fundamentally in scope and purpose. Traditional event extraction methods identify isolated events and their attributes---who, what, when, where---from text \cite{chieu2004query}. INA goes beyond this extraction to \emph{connect events into coherent narratives}, emphasizing not just what happened but how events relate to each other, how they form storylines, and how these storylines evolve and interact over time. Where event mining might produce a database of discrete events, INA produces an interconnected narrative structure that reveals the flow and development of stories.

The distinction between INA and data storytelling or narrative visualization \cite{segel2010narrative,chen2024automation} is particularly important given the overlapping terminology. Data storytelling and narrative visualization focus on \emph{communicating} predetermined insights through carefully crafted narrative techniques---they help authors tell stories with data. In contrast, INA supports \emph{discovering} narratives within data through interactive exploration. This fundamental difference in purpose shapes everything else: data storytelling optimizes for clarity and engagement in presenting known insights, while INA optimizes for exploration and discovery of unknown narrative structures. Similarly, story visualization techniques visualize existing, authored stories such as novels or films where the narrative structure is explicit. INA instead must \emph{extract} implicit narratives from unstructured document collections where no explicit story structure exists, making it a fundamentally different challenge.

\subsection{Core Principles and Objectives}
INA is guided by several core principles and objectives that shape its approach to narrative understanding. At its foundation, INA integrates computational algorithms with human intuition and domain expertise, addressing what Wenskovitch and North \cite{wenskovitch2020interactive} call the "two black-box" problem where both artificial intelligence (AI) systems and human users may lack understanding of each other's reasoning.

INA also strives to balance computational scalability with interactive responsiveness, enabling the processing of large text volumes while supporting real-time user interaction and feedback. This balance is essential for analyzing rapidly evolving news narratives and other high-volume text sources, as noted by Liu et al. \cite{liu2020story} in their work on narrative extraction from breaking news.

Knowledge enhancement represents another crucial principle, as INA incorporates external knowledge resources to improve the accuracy and relevance of narrative extraction. By leveraging domain-specific ontologies, frames, and knowledge graphs, INA systems can produce more contextually appropriate and complete narrative representations, as demonstrated in recent work by Yan and Tang \cite{yan2023narrative} and Blin \cite{blin2022building}.

Supporting narrative sensemaking is the primary objective of INA. Thus, this field seeks to provide tools that help users discover, explore, and understand complex narratives. These tools should enable manipulation and refinement based on user understanding and objectives, creating a more intuitive analytical environment as shown in the work of Keith et al. on semantic interaction for narrative visualization \cite{keith2023iui}.

\begin{figure*}[!htb]
    \centering
    \includegraphics[width=\linewidth]{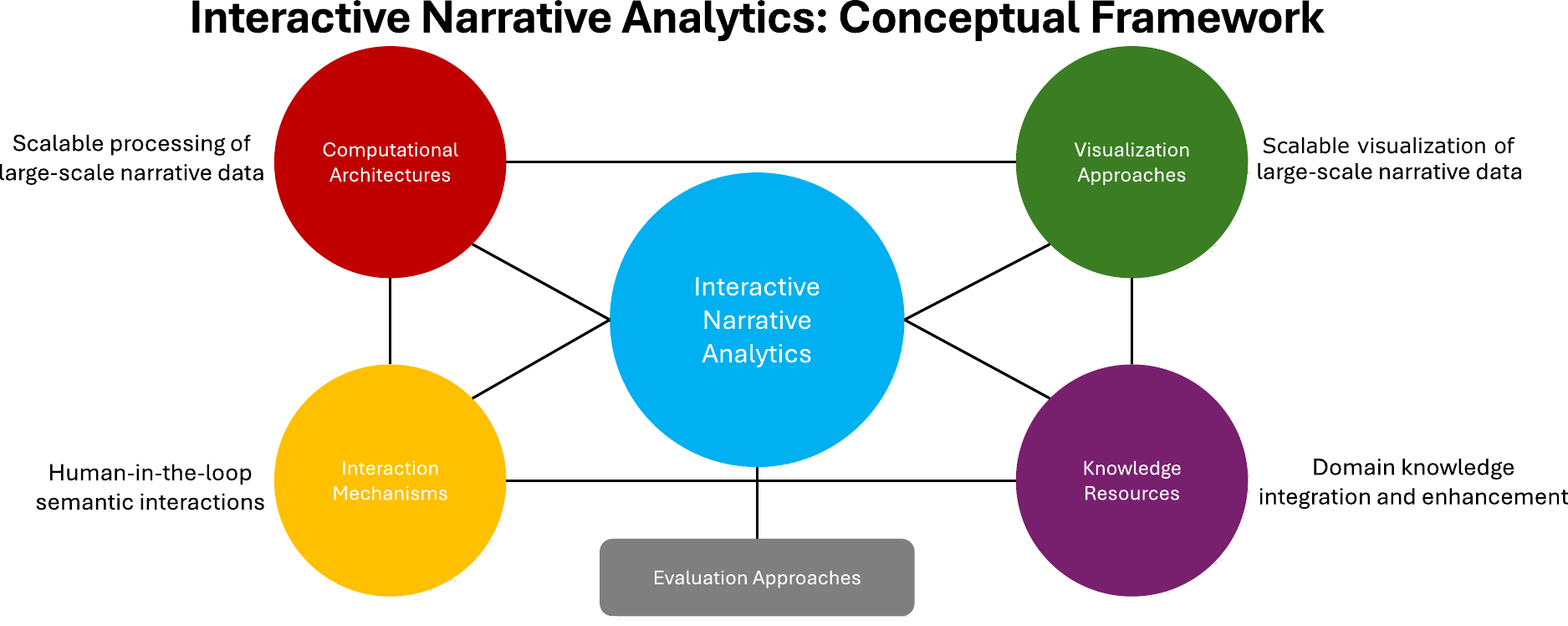}
    \caption{The five core components of Interactive Narrative Analytics and their interconnections. Each component addresses specific challenges while working together in an integrated system.}
    \label{fig:ina-core}
\end{figure*}

Finally, INA emphasizes evaluation and quality assessment through methods for measuring narrative quality, coherence, and utility. These evaluation approaches include metrics for assessing narrative coherence, coverage, and relevance, addressing a significant gap identified in current narrative extraction research \cite{keith2023survey}. Figure \ref{fig:ina-core} presents the five core components of INA.

\begin{table*}[t]
\centering
\caption{Comparative Framework: INA and Related Fields}
\label{tab:comparison}
\resizebox{\linewidth}{!}{%
\begin{tabular}{p{2.2cm}p{2.7cm}p{2.7cm}p{2.7cm}p{2.5cm}p{2.2cm}}
\toprule
\textbf{Aspect} & \textbf{Interactive Narrative Analytics} & \textbf{Visual Analytics} & \textbf{Event Mining} & \textbf{Data Storytelling} & \textbf{Intelligence Analysis} \\
\midrule
\textbf{Primary Goal} & Extract \& explore narrative structures & Analyze complex data patterns & Extract discrete events & Communicate insights & Detect threats \\
\textbf{Input} & Document collections & Any data type & Text corpora & Analyzed data & Multi-source intel \\
\textbf{Output} & Interactive narrative maps & Interactive visualizations & Event databases & Story presentations & Threat assessments \\
\textbf{Temporal Focus} & Central (event sequences) & Variable & Event timestamps & Narrative flow & Real-time monitoring \\
\textbf{User Role} & Explorer/Analyst & Analyst & Consumer & Audience & Intelligence officer \\
\textbf{Interactivity} & Semantic interaction & Data manipulation & Query-based & Limited/guided & Alert-driven \\
\textbf{Knowledge Integration} & Domain ontologies & Statistical models & NLP models & Author knowledge & Classified databases \\
\textbf{Target Scalability} & Thousands of documents with real-time interaction & Varies by data type and visualization & Document-level batch processing & Single dataset/analysis & Real-time streaming with alerts \\
\textbf{Evaluation Metrics} & Narrative coherence, coverage, utility, user task performance & Visual scalability, interaction latency, insight quality & Precision/recall for event extraction & Engagement, comprehension, retention & Detection accuracy, false alarm rate, response time \\
\bottomrule
\end{tabular}%
}
\end{table*}

This comparative framework draws on established characterizations from foundational works and recent surveys across multiple domains. Visual analytics characterizations are based on seminal frameworks by Thomas and Cook \cite{thomas2005illuminating} and Keim et al. \cite{keim2008visual}, with interactive capabilities informed by Endert et al.'s semantic interaction work \cite{endert2011observation,endert2012semantic} and recent advances in large-scale visual analytics \cite{dowling2019interactive}. Event mining approaches reflect methodologies established by Chieu and Lee \cite{chieu2004query} and evolutionary timeline work by Yan et al. \cite{yan2011evolutionary}, while our broader narrative extraction characterization synthesizes comprehensive surveys \cite{keith2023survey,ranade2022computational,santana2023survey}. Data storytelling and narrative visualization draw from Segel and Heer's foundational taxonomy \cite{segel2010narrative} and recent surveys of automation in narrative visualization \cite{chen2024automation}, with visualization scalability considerations informed by information cartography approaches \cite{shahaf2013information}. Intelligence analysis applications reflect the motivations and requirements articulated in Thomas and Cook's work \cite{thomas2005illuminating} and practical visual analytics deployments \cite{shukla2017discrn}. The INA characterization synthesizes our proposed framework with current capabilities in narrative extraction \cite{keith2023survey,keith2020maps,liu2020story}, interactive visualization \cite{endert2011observation,endert2012semantic,keith2023iui}, and knowledge integration \cite{yan2023narrative,blin2022building}. Scalability assessments for each field reflect typical system capabilities and architectural constraints discussed in domain literature, including recent work on parameter-efficient adaptation for large-scale text processing \cite{hu2022lora,dettmers2023qlora}, progressive visualization techniques \cite{ulmer2023survey}, and multi-scale processing systems \cite{cakmak2020multiscale}. Evaluation metrics represent those commonly reported in each field's literature, with narrative quality assessment drawing from recent work on subjective task evaluation \cite{frenda2025perspectivist,leonardelli2023semeval} and established visual analytics evaluation frameworks \cite{keim2008visual}.

\subsection{Integration of Existing Approaches}
The defining attributes of INA work in tandem to enable narrative understanding, building upon and extending established concepts from visual analytics and narrative extraction. The \emph{human-in-the-loop} approach, while present in earlier visual analytics work \cite{endert2011observation,endert2012semantic}, takes on new significance in INA by incorporating user feedback directly into the narrative extraction process rather than merely the visualization layer. This allows algorithms to adapt to user understanding of narrative coherence and importance, which varies by domain and analytical goal.

Semantic interaction techniques in INA extend the foundational work of Endert et al. \cite{endert2011observation,endert2012semantic} by enabling users to manipulate narrative visualizations in ways that reflect their understanding of story structure. When an analyst moves events closer together or groups related storylines, these interactions propagate to the computational model, refining its understanding of narrative relationships. This bidirectional flow between user and system, recently advanced through deep learning approaches \cite{bian2021deepsi,han2023explainable}, creates a more intuitive environment specifically tailored for narrative analysis rather than general data exploration.

Multi-scale representation in narrative contexts presents unique challenges beyond traditional overview-plus-detail visualizations \cite{angelini2018review}. INA must maintain narrative coherence across scales---from individual events to complete storylines to entire narrative landscapes. Building on information cartography concepts \cite{shahaf2013information}, INA systems support fluid transitions between reading individual documents, following specific narrative threads, and understanding the overall narrative structure. This requires careful attention to preserving temporal and causal relationships across scales, ensuring that zooming in or out does not break the narrative flow as the user progressively interacts with the system \cite{fekete2016progressive}.

Knowledge integration in INA goes beyond simple entity linking to incorporate rich domain knowledge about narrative structures, event types, and causal relationships. Recent work on narrative-enhanced knowledge graphs \cite{yan2023narrative,blin2022building} demonstrates how external knowledge can fill gaps in extracted narratives, provide historical context, and validate causal connections. This integration must be carefully balanced---too little leaves narratives incomplete, while too much can overwhelm the extraction process with irrelevant information.

The adaptive extraction algorithms in INA learn not just from user feedback but from the evolving understanding of what constitutes a coherent narrative in specific domains. This adaptation, advocated by Wenskovitch et al. \cite{wenskovitch2021beyond}, enables systems to improve their narrative extraction over time, learning domain-specific patterns of narrative development, common storyline structures, and important narrative elements that might not be explicitly stated in the text.

\section{Theoretical Foundations}
INA draws from four key theoretical traditions that have evolved over decades of research in their respective fields. In particular, INA draws from four key theoretical traditions: interactive visual analytics \cite{wenskovitch2020interactive}, computational narratives \cite{keith2023survey,ranade2022computational,santana2023survey}, sensemaking \cite{pirolli2005sensemaking,kintsch1998comprehension,zwaan1998situation,bruner2009actual}, and knowledge representation \cite{porzel2022narrativizing}. This section provides a focused overview of how these theoretical foundations converge to establish the conceptual framework of INA.

\textbf{Visual Analytics Foundations.}
The theoretical foundations of visual analytics, essential to INA, were established through seminal works that shaped the field. Thomas and Cook's ``Illuminating the Path'' \cite{thomas2005illuminating} provided the foundational definition of visual analytics as ``the science of analytical reasoning facilitated by interactive visual interfaces,'' explicitly connecting it to intelligence analysis needs. This work, emerging from post-9/11 security concerns, established core principles of combining automated analysis with human insight that remain central to INA.

Keim et al.'s framework \cite{keim2008visual} formalized visual analytics as combining ``automated analysis techniques with interactive visualizations for an effective understanding, reasoning and decision making on the basis of very large and complex data sets.'' This definition explicitly acknowledges the ``two black box'' problem that Wenskovitch and North \cite{wenskovitch2020interactive} later address---neither humans nor machines fully understand each other's reasoning processes. Ben Shneiderman's ``eyes have it'' mantra \cite{shneiderman1996eyes}---overview first, zoom and filter, then details on demand---provides an interaction paradigm that INA extends to narrative exploration.

The intelligence analysis roots of visual analytics directly inform INA's development. Pike et al.'s work on the science of interaction \cite{pike2009science} established principles for how analysts make sense of complex information through interactive visualization. The VAST Challenge series, beginning in 2006, has driven innovation in visual analytics for narrative understanding through scenarios involving terrorism, disease outbreaks, and social unrest. These challenges revealed the need for tools that can extract and visualize narrative structures from massive text collections---precisely the gap INA addresses.

\textbf{Narrative Theory and Computational Approaches.}
Narrative theory provides essential concepts of narrative structure---events, entities, temporal and causal relations---that computational narratology has translated into formal models \cite{keith2023survey,ranade2022computational,santana2023survey}. Event-based approaches conceptualize narratives as sequences of connected events, while entity-based approaches focus on characters and relationships. The concept of narrative coherence \cite{shahaf2010connecting, keith2020maps} is central to computational modeling, measured through semantic similarity, temporal consistency, and causal connectedness.

\begin{figure*}[!htb]
    \centering
    \includegraphics[width=\linewidth]{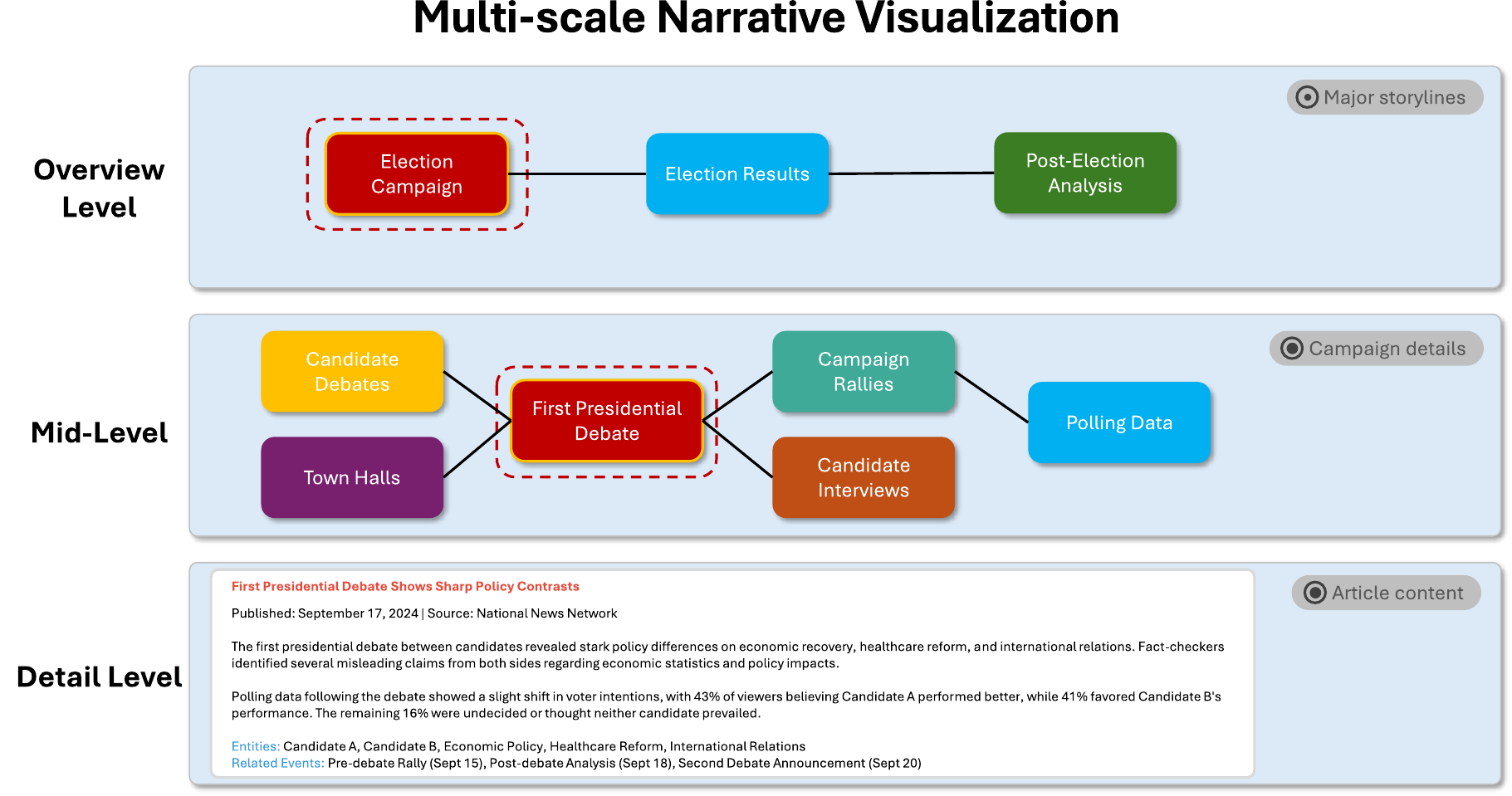}
    \caption{Multi-scale narrative visualization showing different levels of granularity while maintaining context across levels.}
    \label{fig:multiscale}
\end{figure*}

\textbf{Sensemaking.}
Sensemaking, characterized by Pirolli and Card as an iterative process involving information foraging and schema formation, is inherently subjective and context-dependent \cite{pirolli1995information,pirolli2005sensemaking}. Visual analytics \cite{dowling2019interactive} supports this process through interactive visualizations that enable exploration at multiple scales, from overview to detail \cite{shahaf2013information}. Semantic interaction---enabling users to express analytical reasoning through natural interactions that propagate to underlying computational models---is particularly significant for INA \cite{endert2011observation,endert2012semantic,han2023explainable}. Figure \ref{fig:multiscale} shows a mockup of a multiscale INA system.

\begin{figure*}[!htb]
    \centering
    \includegraphics[width=\linewidth]{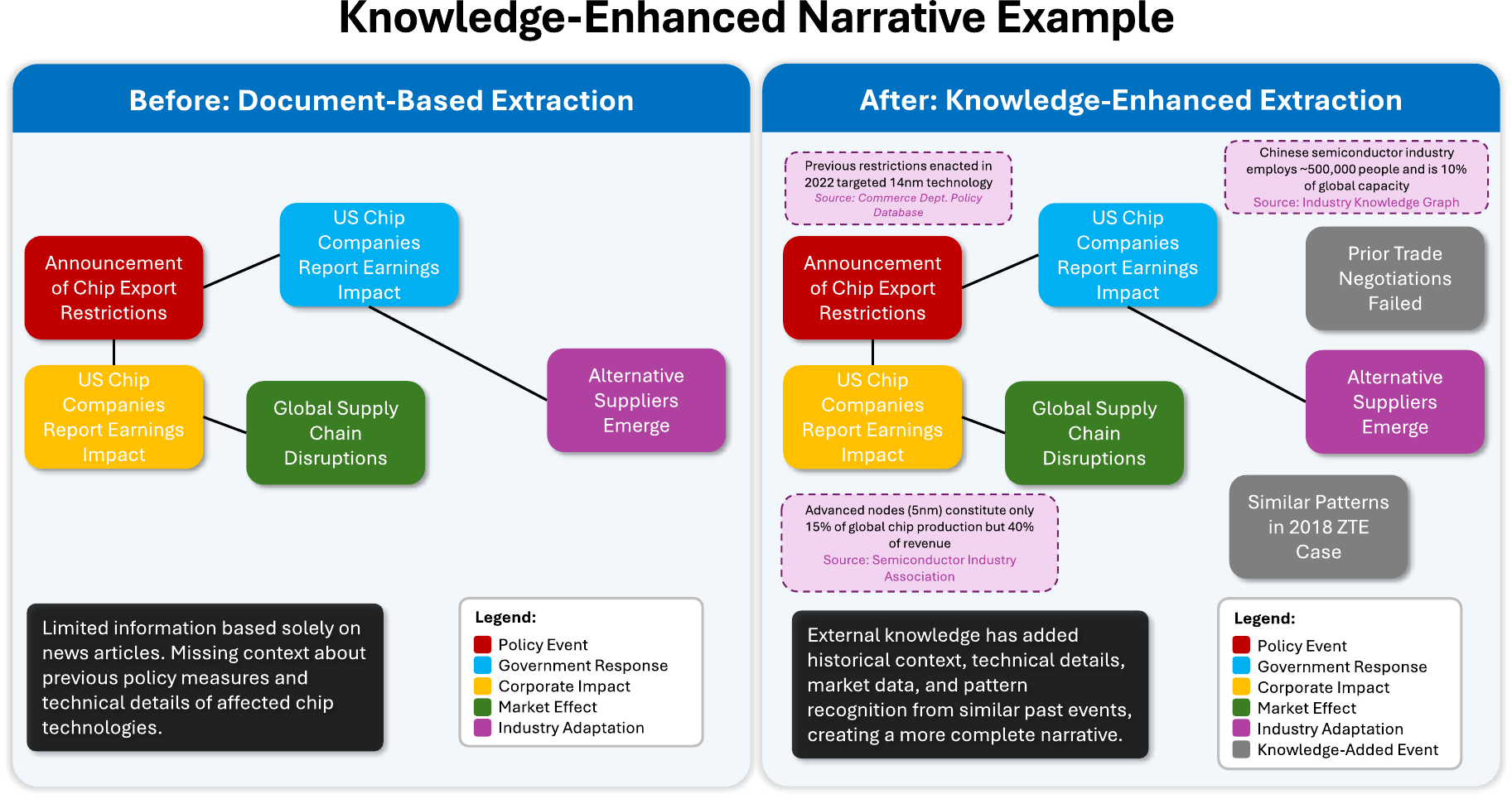}
    \caption{Before-and-after comparison showing how external knowledge enhances a narrative about semiconductor export restrictions with contextual information, historical patterns, and domain expertise.}
    \label{fig:knowledge}
\end{figure*}

\textbf{Knowledge Representation and Integration.}
Knowledge representation provides frameworks for incorporating external knowledge into narrative extraction through ontologies, frames, and knowledge graphs \cite{trojahn2022foundational,yan2023narrative}. Knowledge graphs align naturally with narrative structures \cite{yang2024give}, representing entities, events, and relationships in graph-based formats---though challenges remain in adapting general knowledge resources to specific narrative domains \cite{blin2022building}. Figure \ref{fig:knowledge} shows a representation of how knowledge could augment interactive narrative systems.

\textbf{Convergence in Interactive Narrative Analytics.}
INA represents the convergence of these theoretical traditions, synthesizing concepts from narrative theory, sensemaking, visual analytics, and knowledge representation. This convergence is guided by several key principles:

\begin{itemize}
\item \textbf{Complementarity of Computation and Cognition:} INA recognizes that computational models and human cognition offer complementary strengths in narrative understanding \cite{wenskovitch2020interactive}.

\item \textbf{Iterative Narrative Construction:} INA conceptualizes narrative extraction as an iterative process involving continuous refinement based on user feedback and interaction \cite{wenskovitch2021beyond}.

\item \textbf{Knowledge-Enhanced Understanding:} INA incorporates external knowledge resources to enhance narrative extraction and visualization, recognizing that narrative understanding often requires background knowledge not explicitly stated in text \cite{yan2023narrative}.

\item \textbf{Interactive Exploration:} INA emphasizes interactive exploration in narrative understanding, providing tools that enable users to navigate and manipulate narrative representations according to their analytical goals \cite{keith2023iui}.
\end{itemize}

This theoretical integration provides a foundation for addressing the challenges of narrative sensemaking. Drawing on established theories across multiple disciplines, INA offers a principled approach to combining computational methods with human insight for narrative understanding.

\section{Core Elements: Current Approaches and Research Challenges}
The development of Interactive Narrative Analytics requires advancing five interconnected elements: computational architectures, visualization approaches, interaction mechanisms, knowledge resources, and evaluation frameworks. For each element, we examine current approaches, their limitations, and the research challenges that must be addressed to realize INA's potential.

\subsection{The Need for Interactive Narrative Analytics}
The limitations identified in current approaches highlight the need for an integrated field of INA that addresses these challenges holistically. By combining advances in computational narrative extraction, visual analytics, and interactive systems, INA aims to provide more effective tools for understanding complex narrative landscapes.

Specifically, INA seeks to address scalability challenges through optimized extraction algorithms and efficient visualization techniques, enabling analysis of large-scale narrative collections. It aims to enhance interactivity through semantic interaction models specifically designed for narrative exploration and manipulation, allowing analysts to incorporate their expertise into the extraction process. It strives to integrate external knowledge resources to improve narrative completeness and accuracy, providing context and background information that may be implicit in the text.

In addition, INA strives to develop integrated evaluation frameworks that assess both computational performance and user experience, accounting for the subjective nature of narrative interpretation while providing meaningful quality metrics. It also aims to address misinformation challenges through credibility assessment mechanisms and narrative verification techniques, helping analysts identify and understand misleading content in narrative contexts.

\begin{figure*}[!htb]
    \centering
    \includegraphics[width=\linewidth]{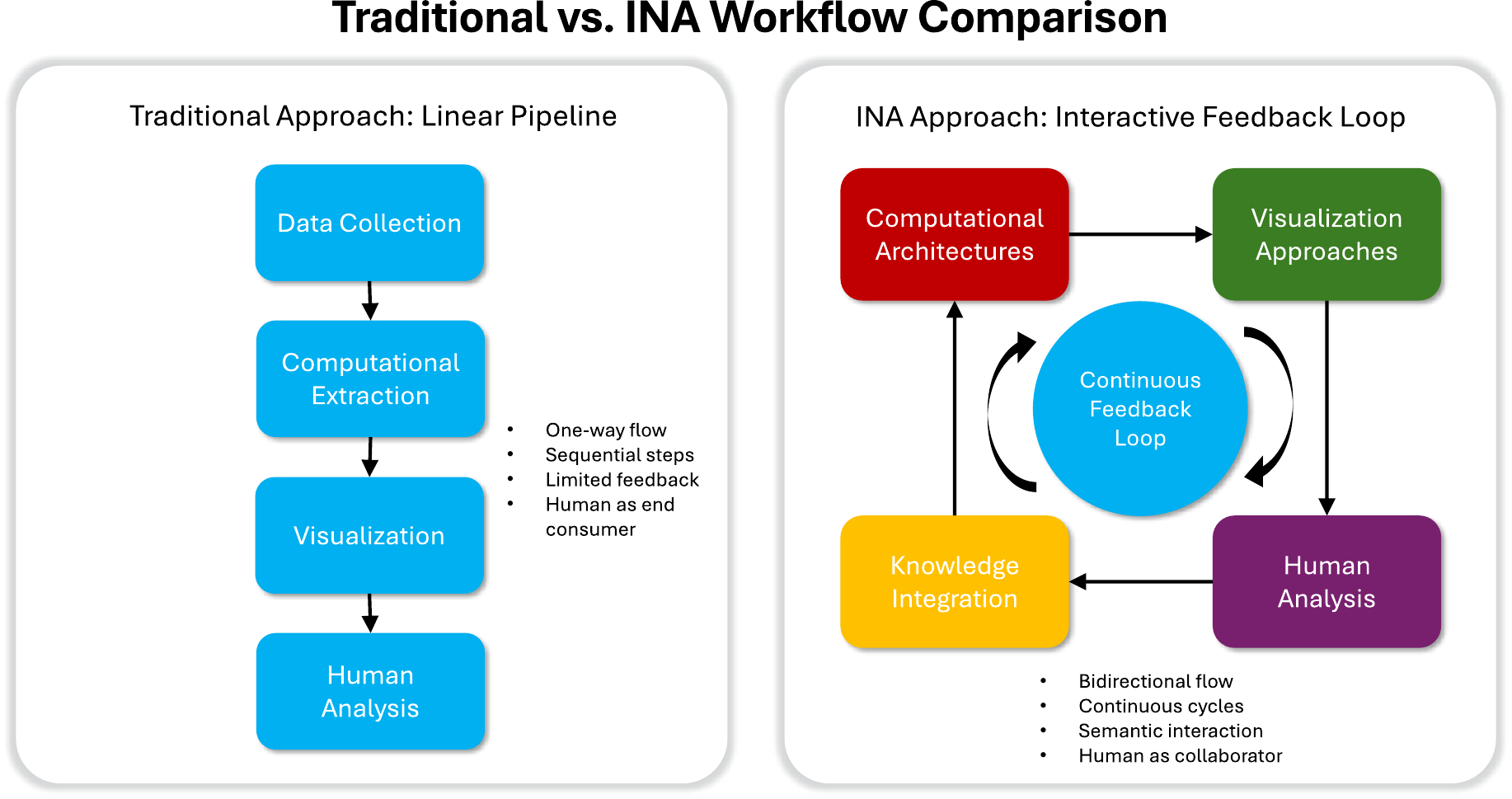}
    \caption{Comparison between the traditional linear pipeline approach and the Interactive Narrative Analytics integrated approach with continuous feedback between computational and human processes.}
    \label{fig:comparison}
\end{figure*}

By addressing these challenges, INA offers a promising approach to narrative understanding in complex information environments, supporting analysts in making sense of the ever-growing volume of textual data in domains ranging from news analysis to intelligence gathering, from scientific literature to social media discourse. Figure \ref{fig:comparison} shows the comparison between INA approaches and traditional pipelines, demonstrating the paradigm shift from sequential processing to integrated, iterative analysis.

\subsection{Computational Architectures for Narrative Processing}
Current narrative extraction methods face fundamental scalability challenges that limit their practical application. Timeline-based approaches like those of Chieu and Lee \cite{chieu2004query} and Yan et al. \cite{yan2011evolutionary} typically exhibit polynomial time complexity in the number of documents, making them impractical for large corpora. More sophisticated timeline extraction by Shahaf and Guestrin \cite{shahaf2010connecting} identifies coherent chains of news articles, yet still faces computational bottlenecks with large collections. Graph-based approaches such as Narrative Maps \cite{keith2020maps} and Story Forest \cite{liu2020story} face even steeper computational challenges, with complexity often quadratic or higher relative to document count. These methods involve computationally intensive operations including pairwise document similarity calculations, global optimization over entire document collections, and complex graph algorithms that become prohibitive as collections grow \cite{keith2023survey}.

The computational demands become particularly acute when narratives evolve in real-time, requiring systems to process streaming data and incrementally update narrative structures without complete recalculation. Current approaches struggle with this temporal constraint, creating tension between computational thoroughness and responsiveness. Shahaf et al. \cite{shahaf2013information} address this through approximation techniques and incremental updates, but these solutions often introduce trade-offs between narrative quality and computational performance that remain unresolved. Most existing systems operate in batch mode, processing entire document collections offline before presenting results, which fails to support the iterative exploration central to INA.

The integration of Large Language Models presents both opportunities and challenges for computational narrative extraction. While LLMs offer powerful semantic processing capabilities that could enhance narrative understanding, they introduce substantial computational costs, potential biases, and transparency issues \cite{choudhary2022interpretation,zhao2024leva, chang2024survey}. Their black-box nature often obscures the reasoning behind extracted narrative structures, making it difficult for analysts to understand how particular narratives were derived. Current research explores using LLMs for specific subtasks like event extraction or coherence scoring, but fully integrating these models into interactive narrative systems while maintaining computational efficiency and interpretability represents a critical research challenge.

Addressing these computational challenges requires developing architectures that balance multiple competing demands. Systems must process document collections ranging from thousands to millions of documents while maintaining sub-second response times for interactive operations. They must support both batch processing for initial narrative extraction and stream processing for real-time updates. They must integrate multiple computational approaches---from traditional NLP to deep learning to graph algorithms---within a unified framework. Recent work on distributed computing frameworks and GPU acceleration shows promise \cite{dowling2019interactive}, but adapting these technologies specifically for narrative processing while maintaining the semantic richness required for meaningful narrative extraction remains an open challenge.

Recent advances in parameter-efficient fine-tuning provide practical pathways for adapting large language models to narrative extraction tasks while maintaining computational feasibility for interactive systems. For example, Low-Rank Adaptation (LoRA) \cite{hu2022lora} decomposes weight updates into low-rank matrices, reducing trainable parameters by orders of magnitude while matching full fine-tuning performance. For resource-constrained environments, QLoRA \cite{dettmers2023qlora} combines 4-bit quantization with LoRA to enable fine-tuning of very large models on single GPUs. Prompt tuning approaches \cite{li2021prefix,lester2021power} offer even greater efficiency by optimizing only continuous task-specific vectors while freezing model parameters entirely. For INA systems, modular integration architectures could allow the selective use of LLMs for specific subtasks (e.g., event extraction, coherence scoring, and entity linking) while maintaining traditional approaches for computationally intensive operations. However, the choice among these techniques involves trade-offs between model size, computational efficiency, adaptation quality, and interactive responsiveness.

Another challenge in this context is addressing the "two black-box" problem, which requires integrating explainable AI techniques specifically adapted for narrative extraction. SHAP-based approaches \cite{mosca2022shap} or LIME variants \cite{ribeiro2016should} could be used to quantify the contribution of individual narrative elements (events, entities, temporal markers) to extraction decisions, though adaptation from tabular to textual narrative data requires careful consideration \cite{keith2025xai}. Counterfactual generation methods \cite{wu2021polyjuice,chen2023disco} could enable analysts to understand how narrative structures would change under alternative event configurations---for instance, how removing specific events would affect inferred causal chains. Attention visualization techniques that go beyond raw attention weights \cite{chefer2021transformer} could be useful to reveal which textual spans most influence narrative connection decisions across transformer layers. For narrative-specific explanations, causal inference approaches \cite{jin2021causal,zhou2023causal} could help surface why particular events are grouped into storylines or why certain temporal/causal relationships are inferred. In this context, developing \textit{interactive explanation interfaces} could allow users to query why specific narrative elements were connected, drill down into model reasoning, and provide corrective feedback that propagates through the extraction model---creating a transparent and collaborative human-in-the-loop system where analysts understand and can guide computational narrative construction.

\subsection{Visualization Approaches for Narrative Structures}
Creating effective visual representations of narrative structures presents unique challenges beyond traditional information visualization. Narratives inherently involve complex structures with multiple storylines, branching paths, and interconnected elements that resist straightforward visual encoding. Current approaches broadly fall into timeline-based, graph-based, and hybrid visualizations, each with distinct capabilities and limitations.

Timeline-based visualizations excel at representing temporal evolution but struggle with complex non-linear narratives. Systems like CloudLines \cite{krstajic2011cloudlines} address scalability through compact visualization techniques using logarithmic time scaling and distortion lenses, yet they primarily support linear narrative flows. Shahaf et al.'s Information Cartography \cite{shahaf2013information} extends timeline concepts with zoomable maps supporting multiscale exploration, but maintaining narrative coherence across scales remains challenging. The fundamental limitation of timeline approaches lies in their difficulty representing parallel storylines, narrative branches, and convergent paths that characterize complex real-world narratives \cite{keith2023survey}.

Graph-based visualizations offer greater expressiveness for complex narrative structures but introduce severe scalability and interpretability challenges. As narrative graphs grow, they quickly become cluttered and difficult to navigate \cite{dowling2019interactive}. Keith and Mitra's Narrative Maps \cite{keith2020maps} attempt to address this through hierarchical clustering and progressive disclosure, but users still struggle to maintain awareness of overall narrative structure while examining specific details. Liu et al. \cite{liu2020story} organize events into tree structures that capture evolutionary relationships, yet these still face visualization challenges as trees grow. The cognitive load of interpreting complex graph structures can overwhelm analysts, particularly when narratives involve hundreds of events and multiple intersecting storylines. Furthermore, standard graph layout algorithms often fail to preserve the temporal and causal relationships crucial to narrative understanding.

\begin{figure*}[!htb]
    \centering
    \includegraphics[width=\linewidth]{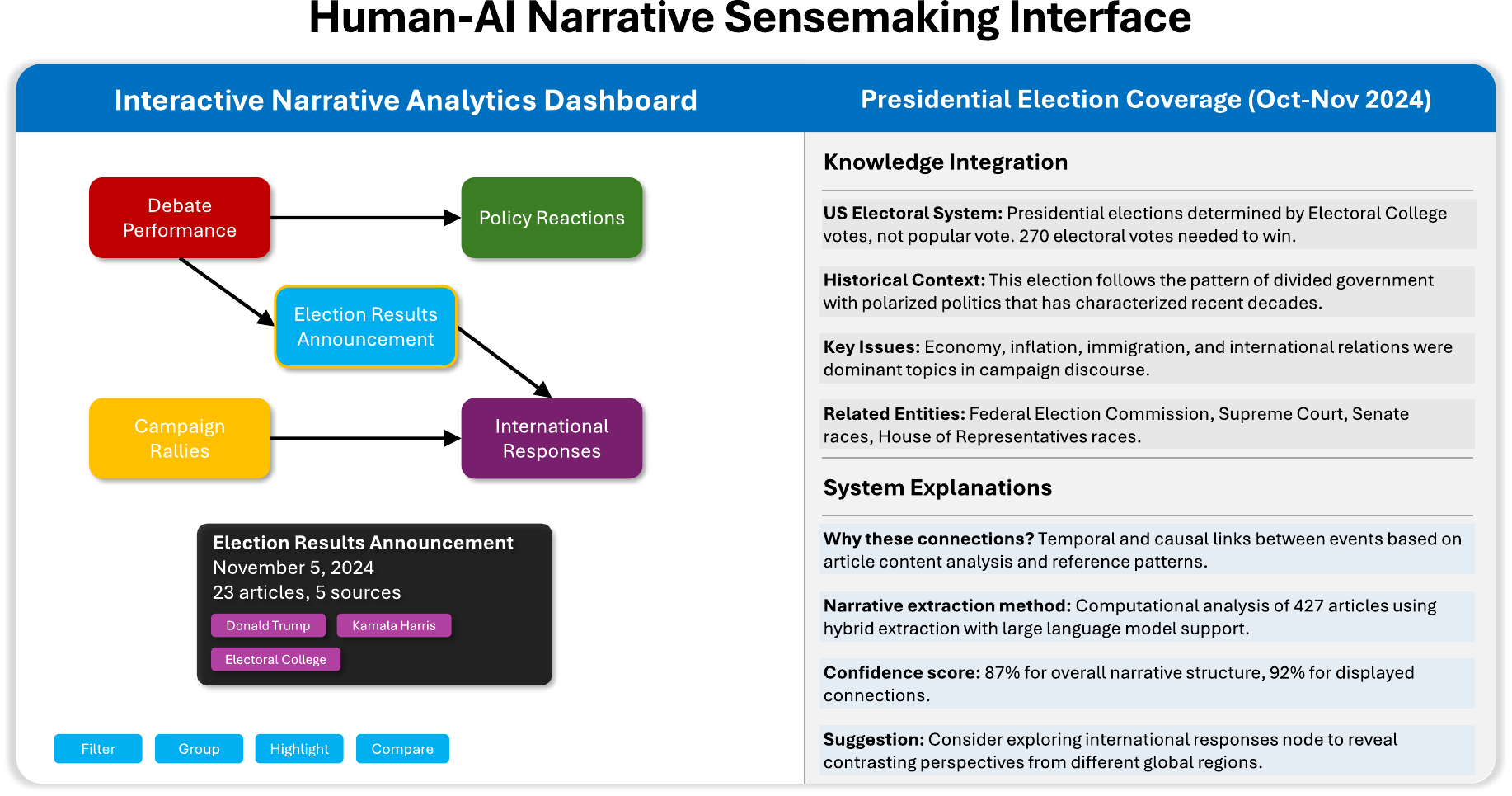}
    \caption{Mockup of an Interactive Narrative Analytics interface showing a narrative map visualization with semantic interaction capabilities and knowledge integration features.}
    \label{fig:mockup}
\end{figure*}

The design of these visualizations must consider human perceptual and cognitive constraints \cite{johnson1983mental, repovvs2006multi}. Limited working memory, attentional bottlenecks, and interpretation biases all affect how effectively analysts can make sense of visual narrative representations \cite{hasher1988working,slagter2007mental,fair2007development}. Visualization approaches must work within these constraints while still conveying the richness and complexity of narrative structures. Figure \ref{fig:mockup} shows a mockup of a potential INA system that addresses these challenges through multi-level representation, semantic interaction capabilities, and integrated knowledge features.

Hybrid approaches attempt to combine the strengths of timeline and graph representations. Angus et al.'s Discourse Lines \cite{angus2023discourse} blend timeline representations with narrative maps to show policy and media storylines, using spatial positioning to encode both temporal progression and thematic relationships. These hybrid visualizations show promise but introduce new challenges in maintaining visual coherence, managing screen real estate, and helping users understand multiple visual encodings simultaneously. The design space for hybrid narrative visualizations remains largely unexplored, with no established guidelines for when to use which visual encoding or how to transition between them \cite{keith2023iui}.

Topic and trend visualization techniques, while not specifically designed for narratives, provide relevant insights for INA design. Topic models visualize thematic patterns in text collections, though they typically operate at document level rather than capturing fine-grained event structures \cite{dowling2019interactive}. Trend visualization shows temporal patterns but often lacks the causal and explanatory connections between events that define narratives \cite{ranade2022computational}. These approaches highlight the gap between general text visualization and narrative-specific visual representations.

In this context, several multi-scale techniques could address some of the visualization challenges inherent in complex narrative structures. Edge bundling methods \cite{wallinger2021edge,lhuillier2017state} reduce visual clutter in narrative graphs by routing related connections along shared paths while maintaining readability, with recent optimizations achieving significant speedup for large-scale graphs. Timeline compression approaches enable semantic zoom where temporal granularity adjusts based on user focus---for instance, showing individual events during detailed examination but aggregating to weekly or monthly summaries in overview mode \cite{cakmak2020multiscale}. Hierarchical clustering techniques organize narratives into multi-level structures allowing top-down exploration from major storylines to constituent events \cite{yang2020interactive}, with progressive visualization methods \cite{ulmer2023survey} enabling incremental refinement as users navigate across scales. The key challenge in using these approaches lies in maintaining narrative coherence across scale transitions---ensuring that temporal ordering, causal relationships, and thematic connections remain interpretable whether viewing individual documents, specific narrative threads, or the entire narrative landscape.

\subsection{Interaction Mechanisms for Narrative Exploration}
The interactive dimension fundamentally distinguishes INA from automated narrative extraction, yet current systems provide limited support for meaningful interaction with narrative structures. Most existing narrative extraction methods operate in batch mode with minimal user involvement \cite{keith2023iui}, while those that do support interaction typically limit it to basic operations like filtering, zooming, and highlighting. Search and filtering interfaces rely heavily on users formulating appropriate queries and synthesizing results into coherent narratives themselves \cite{dowling2019interactive}. This gap between automated extraction and interactive exploration represents a critical limitation in current approaches.

Semantic interaction, pioneered by Endert et al. \cite{endert2011observation,endert2012semantic} and extended through deep learning by Bian and North \cite{bian2021deepsi}, enables analysts to express analytical reasoning through natural interactions with visualizations. When users spatially arrange documents or adjust similarities, these interactions propagate to underlying computational models. Recent advances by Han et al. \cite{han2023explainable} demonstrate how explainable projections can make these interactions more interpretable. However, current semantic interaction systems focus primarily on document-level representations rather than narrative structures \cite{keith2023survey}. Extending semantic interaction to narrative elements---events, entities, relationships, and storylines---requires new interaction paradigms that can capture the complex reasoning involved in narrative sensemaking.

Keith et al. \cite{keith2023iui} introduce mixed multi-model semantic interaction specifically for narrative visualizations, allowing users to manipulate graph-based narrative representations with changes propagating through multiple computational models. Their approach demonstrates how user interactions with visual narrative elements can update both the extraction model's understanding of narrative coherence and the relevance weights for different narrative components. Yet this work also reveals fundamental challenges: inferring user intent from potentially ambiguous interactions, translating those intentions into meaningful model updates, and managing conflicts between user preferences and computational recommendations all remain difficult problems.

The challenge of interaction design for narrative analytics extends beyond technical implementation to understanding how analysts naturally think about and work with narratives. Cognitive processes involved in narrative sensemaking \cite{lerner2011topographic,bietti2019storytelling} include identifying key events, establishing causal connections, recognizing patterns across storylines, and synthesizing multiple perspectives into coherent understanding. Current interaction mechanisms often fail to support these natural reasoning processes, forcing analysts to translate their narrative understanding into system-specific operations. Developing interaction vocabularies that align with analysts' mental models of narratives while remaining computationally tractable represents an ongoing challenge \cite{endert2011observation,endert2012semantic}.

System responsiveness creates additional technical challenges when interactions trigger computationally intensive operations. Re-extracting narratives based on user feedback, updating visualizations to reflect new parameters, or computing alternative narrative structures all potentially require significant computation. Current systems often sacrifice either interactivity (by processing updates offline) or quality (by using simplified models for real-time response). Progressive computation approaches that provide initial results quickly while refining them over time show promise \cite{stolper2014progressive}, but implementing these approaches specifically for narrative analytics while maintaining narrative coherence requires careful algorithm design and avoiding the issue of introducing potentially spurious patterns \cite{zgraggen2018investigating}.

Multimodal interaction modalities could enhance narrative exploration through natural, accessible interfaces. For example, voice command interfaces could enable hands-free navigation of narrative structures, particularly valuable for large display environments where analysts may be distant from traditional input devices \cite{leon2024talk}. Natural language query systems allow analysts to ask questions like "show me events related to trade negotiations in March" rather than constructing complex filter sequences \cite{shen2022towards}. Gesture control for spatial narrative manipulation enables direct manipulation of graph-based narrative visualizations through touch or mid-air gestures, creating more intuitive workflows for exploring connections between events. Eye-tracking systems could implicitly detect analyst interest, automatically providing additional detail for fixated elements or adapting visualizations based on attention patterns. However, such multimodal interaction capabilities remain unexplored in INA and could present interesting opportunities for future work in the field.

\subsection{Knowledge Resources for Narrative Enhancement}
External knowledge integration fundamentally enhances narrative extraction by providing context, background information, and domain expertise that may not be explicit in text. However, incorporating knowledge resources into narrative analytics systems presents significant challenges in representation, alignment, and reasoning that current approaches only partially address.

Knowledge representation for narratives must balance expressive power with computational feasibility. Ontologies and frames can capture rich narrative concepts---event types, participant roles, causal relationships---but often become unwieldy for large-scale processing. Knowledge graphs offer more scalable representations and align naturally with graph-based narrative structures \cite{yan2023narrative, blin2022building}, yet they struggle to represent the temporal dynamics and uncertainty inherent in evolving narratives. Porzel et al. \cite{porzel2022narrativizing} explore narrativizing knowledge graphs to better capture story-like structures, but finding representations that can capture narrative complexity while supporting efficient computation remains challenging.

The alignment between textual mentions and knowledge resources introduces fundamental difficulties in narrative contexts. Entity linking in narratives must handle not just ambiguity and coreference but also temporal evolution---entities change over time, relationships shift, and the same entity may play different narrative roles in different contexts. Event detection and alignment face even greater challenges, as the same real-world event may be described differently across sources, at different granularities, and from different perspectives. Current approaches using neural methods show promise for entity and event alignment, but they often fail to capture the narrative-specific aspects of these problems, such as maintaining character continuity or preserving causal chains \cite{yan2023narrative}.

Domain adaptation presents another critical challenge for knowledge-enhanced narrative analytics. Different narrative domains---news, scientific literature, legal documents---require different knowledge resources and reasoning patterns. A system analyzing news narratives needs knowledge about current events, political entities, and geographic relationships, while scientific narrative analysis requires understanding of research concepts, methodologies, and citation patterns. Developing domain-specific knowledge resources requires substantial expertise and effort, yet general-purpose resources often lack the specificity needed for meaningful narrative enhancement \cite{blin2022building}. Transfer learning and few-shot adaptation techniques offer potential solutions, but their application to narrative-specific knowledge remains largely unexplored.

Handling noisy and conflicting knowledge graph data requires robust mechanisms at multiple levels \cite{liang2024survey,deng2023gold}. For instance, confidence scoring systems could weight knowledge sources based on provenance, recency, and historical reliability, allowing narrative extraction to prioritize high-quality information while still considering alternative perspectives. Provenance tracking \cite{sikos2020provenance} mechanisms must maintain detailed records of where knowledge originated, enabling analysts to trace claims back to sources and assess credibility in context. In this context, conflict resolution heuristics become necessary when multiple sources provide contradictory information about events or relationships: voting-based approaches can favor majority consensus, recency-based methods can prioritize newer information for evolving situations, and source reliability-based techniques can defer to authoritative sources.

The integration of multiple knowledge sources compounds these challenges. Real-world narrative analysis often requires combining general knowledge (from resources like WikiData or ConceptNet) with domain-specific knowledge (from specialized ontologies) and temporal knowledge (from event databases or news archives). These resources may use different representations, cover different time periods, and contain conflicting information. Reconciling these differences while maintaining narrative coherence requires sophisticated fusion techniques that can reason about source reliability, temporal validity, and contextual relevance \cite{porzel2022narrativizing}.

\subsection{Evaluation Approaches for Narrative Quality}
The assessment of INA systems presents distinctive challenges related to the complex, subjective nature of narrative understanding. Developing metrics and methodologies that meaningfully capture narrative quality, system usability, and analytical utility remains an open research problem \cite{keith2023survey}. The lack of standardized evaluation frameworks significantly hinders comparative assessment and systematic progress in the field.

Narrative quality assessment involves multiple dimensions including coherence, completeness, accuracy, and relevance. Formalizing these subjective qualities into measurable properties requires careful consideration of narrative theory, computational feasibility, and human judgment alignment. Different analytical contexts may prioritize different quality aspects, further complicating standardization efforts. Current metrics often capture isolated aspects of narrative quality but fail to provide comprehensive assessment aligned with human narrative perception.

User-centered evaluation introduces additional complexity, as it must assess how effectively systems support analytical tasks within realistic contexts. Methodologies must balance experimental control with ecological validity, requiring careful task design and appropriate measurement approaches. The diverse analytical needs across different application domains further complicate standardization efforts. Developing formal evaluation approaches that are also practical represents a significant challenge for advancing INA research. Future empirical studies comparing INA approaches should carefully consider experimental design choices---such as between-subjects versus within-subjects comparisons---and ensure sufficient statistical power to detect meaningful differences in user performance and narrative quality.

Comparative evaluation across different INA systems requires standardized datasets, tasks, and metrics that remain largely undeveloped. Creating representative benchmark materials that capture diverse narrative phenomena while supporting systematic comparison requires substantial community effort. The subjective nature of narrative interpretation further complicates ground truth establishment, as multiple valid interpretations may exist for the same narrative material. Addressing these evaluation challenges will require collaborative efforts across the research community and integration of perspectives from multiple disciplines.

In particular, addressing subjectivity in narrative interpretation requires evaluation frameworks that embrace rather than eliminate disagreement \cite{leonardelli2023semeval}. Thus, annotation protocols that preserve individual annotator judgments---rather than simply collapsing to a majority vote---are required to retain valuable disagreement signals, as perfect agreement may be impossible---and even undesirable---for subjective narrative tasks \cite{frenda2025perspectivist}. Multi-annotator modeling approaches \cite{fornaciari2021beyond} treat disagreement as informative signal, predicting distributions over possible interpretations rather than single labels, with evaluation metrics that reward capturing this diversity. Task-based evaluation that measures utility for downstream analytical goals (e.g., successful investigation completion, accurate threat assessment) can complement agreement-based metrics, assessing whether systems enable effective sensemaking regardless of "ground truth" \cite{rottger2022two}. Importantly, evaluation must recognize that disagreement sometimes reflects systematic differences across demographic or ideological groups \cite{fleisig2023majority} rather than random noise, requiring fairness metrics that ensure minority perspectives receive appropriate representation in both extraction and evaluation.

Ethical considerations in evaluation add another layer of complexity. Evaluation must consider not just technical performance but also potential biases, privacy implications, and societal impacts \cite{friedman2021eight, chang2024survey}. Systems that excel at extracting narratives might also be effective at surveillance or spreading misinformation \cite{lazer2018science, vosoughi2018spread, menczer2020attention}. Evaluation frameworks must therefore incorporate ethical dimensions, assessing whether systems respect privacy, avoid amplifying biases, and prevent misuse. The challenge of detecting and mitigating misinformation within narrative structures requires specific evaluation criteria \cite{aimeur2023fake, sharma2019combating}, as systems must help analysts identify potentially misleading content while understanding how false narratives spread and evolve.

\subsection{Ethical Considerations in Narrative Analytics}
The development and deployment of INA systems raise critical ethical considerations that must be addressed as the field matures. Fischer et al.'s work on ethical awareness in communication analysis \cite{fischer2022promoting} provides a framework particularly relevant to INA applications. Their MULTI-CASE framework \cite{fischer2024multicase} demonstrates how transformer-based systems can incorporate ethical considerations directly into the analytical process, implementing equal joint agency between human and AI for heterogeneous information analysis.

Privacy concerns in narrative extraction are paramount, as personal narratives extracted from social media, news, or other sources may contain highly sensitive information about individuals and communities. The aggregation of narrative fragments across multiple sources can reveal patterns and connections that individuals never intended to make public. INA systems must implement appropriate consent mechanisms, data minimization principles, and purpose limitation as required by regulations like GDPR. The challenge intensifies when narratives span multiple jurisdictions with different privacy regulations and cultural expectations about information sharing.

Algorithmic bias in narrative interpretation presents another critical challenge. Narrative understanding is inherently cultural, and systems trained on Western news sources may misinterpret narratives from other cultural contexts. The selection of which events to include, how to connect them, and what constitutes narrative coherence all embed cultural assumptions that may not transfer across contexts. Addressing this requires diverse training datasets, regular auditing of system outputs for bias, and inclusive design processes that involve stakeholders from multiple cultural backgrounds.

The potential for INA systems to be used for surveillance or population control cannot be ignored. The same capabilities that help journalists understand complex stories or researchers track scientific developments could be used to monitor dissent, track activist movements, or suppress minority narratives. Developers must consider implementing technical safeguards, use restrictions, and transparency measures that prevent misuse while enabling legitimate applications. This includes careful consideration of who has access to these tools and for what purposes.

Thus, considering the previous concerns, operationalizing ethical INA systems requires concrete technical safeguards and governance mechanisms. Bias auditing pipelines should systematically evaluate narrative extraction across demographic groups, geographic regions, and ideological perspectives \cite{gallegos2024bias}, testing whether certain communities' narratives are systematically underrepresented or mischaracterized. Fairness-aware modeling techniques must ensure balanced narrative extraction across diverse sources, avoiding over-reliance on dominant perspectives while properly weighting minority viewpoints. Privacy-preserving computation becomes critical when narratives involve sensitive information about individuals or communities; differential privacy approaches \cite{fu2024differentially} and federated learning can enable aggregate narrative analysis while protecting individual data points. Transparency measures should include model cards \cite{mitchell2019model} documenting intended use cases, known limitations, and performance across demographic groups, alongside datasheets \cite{gebru2021datasheets} specifying data collection processes and potential biases. Governance frameworks must establish oversight mechanisms for INA deployment, particularly in high-stakes contexts like intelligence analysis or misinformation detection, ensuring that systems remain accountable and contestable rather than becoming opaque arbiters of narrative truth.

\section{Future Research Directions}
\textbf{Advanced Computational Models for Narrative Understanding.}
Future research should focus on developing more powerful and efficient computational models for narrative extraction and understanding. Narrative models based on LLMs offer significant potential but require adaptation for narrative-specific tasks \cite{zhao2024leva, chang2024survey}. Research priorities include developing specialized architectures or training objectives that capture narrative structures, creating more efficient models suitable for interactive applications, and enhancing model interpretability to support human-AI collaboration \cite{wilson2018collaborative,vaccaro2024combinations}. Incremental and adaptive extraction algorithms represent another critical research direction, enabling systems to efficiently update narrative structures as new information becomes available rather than reprocessing entire document collections \cite{liu2020story}. These approaches must address challenges in maintaining narrative coherence across updates while ensuring computational efficiency suitable for interactive exploration.

\textbf{Human-AI Collaboration for Narrative Sensemaking.}
Effective integration of human intelligence and computational capabilities represents a fundamental research direction for INA \cite{wenskovitch2020interactive, wenskovitch2021beyond}. This research should explore different models for human-AI collaboration in narrative understanding, examining how to allocate responsibilities between analysts and algorithms to leverage their complementary strengths. Key research questions include developing appropriate division of cognitive labor, designing interaction mechanisms that support smooth transitions of control, and creating collaborative frameworks that enhance rather than diminish human analytical capabilities. These questions require interdisciplinary approaches that combine insights from human-computer interaction, cognitive science, and AI.

\textbf{Knowledge-Enhanced Narrative Analytics.}
The integration of external knowledge into narrative extraction presents significant opportunities for enhancing the completeness, accuracy, and context of extracted narratives \cite{blin2022building, yan2023narrative, porzel2022narrativizing}. Future research should develop approaches for effectively incorporating domain-specific knowledge and commonsense reasoning into narrative analytics systems.  Research priorities include developing domain-specific knowledge resources for narrative domains, creating efficient alignment techniques for mapping textual mentions to knowledge bases, and designing inference mechanisms that apply external knowledge to enhance narrative understanding. This integration must balance knowledge depth with computational efficiency to maintain interactive performance.

\textbf{Evaluation Frameworks and Ethical Considerations.}
The development of standardized evaluation frameworks represents a critical requirement for advancing INA research. These frameworks should include narrative quality metrics that formalize subjective aspects of narrative coherence and utility, benchmark datasets that enable systematic comparison between approaches, and user-centered evaluation methodologies that assess how effectively systems support analytical tasks \cite{keith2020maps}. In addition to technical development, research must address ethical considerations such as detection and mitigation of bias, protection of privacy in narrative analysis, and transparency improvements that make extraction processes more understandable for users \cite{chang2024survey, friedman2021eight}.

\textbf{Anticipated Limitations for INA Systems.} 
In summary, while INA offers promising approaches to narrative understanding, several fundamental limitations must be acknowledged. Computational complexity presents a persistent challenge: extracting coherent narratives from massive document collections while maintaining interactive responsiveness requires careful algorithmic design and potentially costly infrastructure. The subjective nature of narrative interpretation means that "correct" narrative extraction may not exist---different analysts with different goals may legitimately construct different narrative representations from the same documents. Domain transfer remains difficult, as narrative extraction systems trained on news data may perform poorly on scientific literature or legal documents without substantial adaptation. Evaluation standardization proves challenging given the lack of objective ground truth for many narrative sensemaking tasks, requiring development of task-based metrics that measure analytical utility rather than algorithmic accuracy. Finally, misinformation and adversarial narratives pose unique challenges, as malicious actors may deliberately construct misleading narrative structures that exploit system assumptions about coherence and plausibility.

\section{Applications and Opportunities}
\textbf{News Analysis and Misinformation Detection.}
News analysis represents a primary application domain for INA, addressing challenges in tracking evolving stories and identifying misinformation. INA systems can extract and visualize narrative structures from diverse sources, revealing how stories develop over time and how different publications characterize events \cite{liu2020story}. These capabilities support comparative analyses, identification of underreported aspects, and contextual verification of claims. In misinformation detection, INA can provide contextual verification mechanisms that place specific claims within broader narrative structures \cite{aimeur2023fake, sharma2019combating}. In particular, social media platforms generate vast amounts of content with complex diffusion patterns and embedded misinformation. INA can help track how narratives spread and evolve across platforms and communities, revealing adaptation patterns and transformations \cite{ansah2019graph, liu2017growing}.

In particular, INA systems could integrate multiple complementary components into a unified workflow. Recent work demonstrates the importance of narrative structure analysis for identifying coordinated disinformation campaigns \cite{nandi2025psychology,sosnowski2025dinam}. An INA-based misinformation detection system would combine: \textbf{(1)} stance detection using retrieval-augmented approaches to identify how claims propagate across narratives \cite{zhu2025ratsd}, \textbf{(2)} fact-checking knowledge base integration drawing from datasets like AVeriTeC \cite{schlichtkrull2023averitec} and FEVER \cite{thorne2018fever} for evidence-based verification, \textbf{(3)} source credibility assessment \cite{cosentino2025source} tracking provenance and historical reliability patterns within narrative contexts, \textbf{(4)} inconsistency detection \cite{wu2023mfir} identifying contradictions in evolving narratives over time, and \textbf{(5)} narrative-based claim verification that examines not just isolated statements but their embedding within broader story structures \cite{sosnowski2025dinam}. This integrated approach addresses the limitation that misinformation often manifests through narrative framing and psychological mechanisms \cite{ganti2023narrative} rather than simple factual errors, requiring systems that can surface how stories evolve and how different sources characterize events.

\textbf{Intelligence Analysis and Security Applications.}
Intelligence analysis involves making sense of complex, incomplete, and sometimes contradictory information. INA can assist by extracting narrative elements from intelligence reports and visualizing connections between them \cite{shukla2017discrn}. These capabilities help analysts discover non-obvious relationships between events, identify gaps in intelligence coverage, and construct a more thorough understanding of complex situations. INA systems can also support the generation and testing of hypotheses by enabling the exploration of alternative narrative explanations, comparing these hypotheses with available evidence \cite{pirolli1995information,pirolli2005sensemaking}.

In particular, building on the visual analytics frameworks that emerged from post-9/11 intelligence needs \cite{thomas2005illuminating}, an INA-based intelligence system would integrate: \textbf{(1)} multi-source narrative fusion combining structured intelligence reports, intercepted communications, open-source news, and social media into unified narrative representations that reveal connections across disparate information streams, \textbf{(2)} temporal pattern detection identifying recurring narrative structures that may signal coordinated activities or evolving threats \cite{cakmak2020multiscale}, \textbf{(3)} anomaly identification highlighting events or relationships that deviate from expected narrative patterns within specific threat contexts, \textbf{(4)} hypothesis generation and testing enabling analysts to construct alternative narrative explanations and evaluate them against available evidence \cite{pirolli1995information,pirolli2005sensemaking}, and \textbf{(5)} multi-scale exploration allowing seamless navigation from strategic overviews of entire situations to tactical details of specific events or actors \cite{ulmer2023survey}. Knowledge integration with threat databases, historical case studies, and domain expertise could enhance narrative extraction while privacy-preserving techniques \cite{fu2024differentially} could enable collaborative analysis across organizational boundaries without compromising sensitive sources.

\textbf{Scientific Literature Exploration.}
The exponential growth of scientific literature creates challenges for researchers attempting to understand research landscapes and identify relevant prior work. INA can help researchers map these landscapes by extracting and visualizing narrative structures from scientific publications \cite{shahaf2013information}. INA systems can also support the identification of literature gaps by revealing underexplored connections or questions within research narratives, and scientific claim verification by tracking how findings evolve through subsequent studies \cite{yang2024give}.

In particular, a scientific literature INA system would combine: \textbf{(1)} research trajectory mapping that traces how specific research questions, methodologies, or findings develop across publications and citation networks, revealing the narrative arc of scientific inquiry \cite{shahaf2013information}, \textbf{(2)} literature gap identification by analyzing narrative structures to reveal under-explored connections, contradictory findings that lack resolution, or logical next steps in research progressions, \textbf{(3)} claim verification across studies tracking how specific scientific claims are supported, refuted, or refined through subsequent research \cite{schlichtkrull2023averitec}, enabling researchers to understand evidential strength and ongoing controversies, \textbf{(4)} cross-disciplinary connection discovery identifying narrative threads that span traditional disciplinary boundaries, revealing potential synergies or common patterns across fields, and \textbf{(5)} methodological evolution tracking showing how experimental approaches, analytical techniques, or theoretical frameworks develop and spread through research communities, while acknowledging that scientific narrative interpretation may vary across theoretical traditions or methodological schools \cite{frenda2025perspectivist}.

\section{Conclusions}
We have defined \textbf{Interactive Narrative Analytics} as a nascent field addressing the challenges of understanding complex narrative structures in our information-rich landscape. Rather than treating narrative extraction and interactive visualization separately, \textbf{we establish INA as a distinct interdisciplinary field that integrates these approaches to solve challenges neither computational nor manual methods can address alone}. Furthermore, current narrative analysis suffers from fundamental limitations: computational methods operate as black boxes, visualization systems lack narrative-specific capabilities, and both fail to incorporate external knowledge effectively. These limitations create a significant gap between theoretical potential and practical utility.

INA represents a paradigm shift by \textbf{placing human-machine collaboration at the center of narrative sensemaking}, challenging the notion that extraction should merely precede human analysis. Our five core components---scalable computational architectures, interactive visualization techniques, semantic interaction models, knowledge integration approaches, and evaluation metrics---define the scope of this field.

Finally, as information volumes grow and narrative manipulation becomes more sophisticated, INA offers a \textbf{fundamental reconceptualization of how research communities approach complex narrative landscapes}. We invite researchers across disciplines to contribute to developing this field's foundations, methodologies, and applications.

\section*{Acknowledgment}
This research is funded by the ANID FONDECYT 11250039 Project. The author is also supported by Project 202311010033-VRIDT-UCN.

\bibliographystyle{IEEEtran}
\bibliography{bibliography}

\begin{IEEEbiography}[{\includegraphics[width=1in,height=1.25in,clip,keepaspectratio]{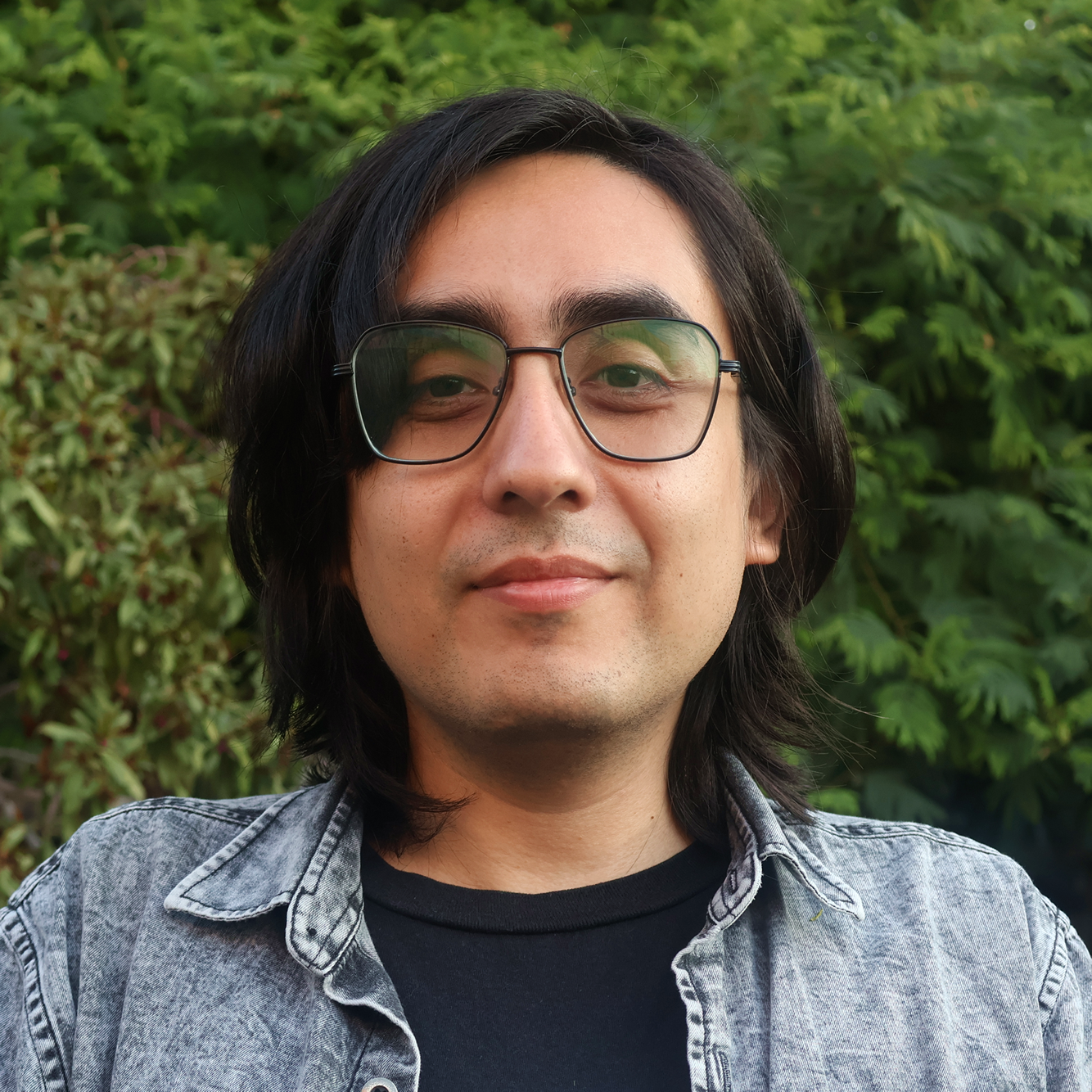}}]{Brian Keith} 
received the B.Sc. degree in engineering and the professional title in computing and informatics civil engineering from Universidad Católica del Norte (UCN), Antofagasta, Chile (2016), the B.Sc. degree in mathematics and the M.Sc. degree in informatics engineering at UCN (2017), and the Ph.D. degree in computer science and applications from Virginia Tech, Blacksburg, VA, USA (2023). He is currently an Assistant Professor with the Department of Systems and Computing Engineering and Secretary of Research and Technological Development for the Faculty of Engineering and Geological Sciences at UCN. He is also the Director of the Artificial Intelligence Innovation Center for the Antofagasta Region (CIARA). He is the author of more than 60 research articles. His research interests include visual analytics, artificial intelligence, text analytics, computational narratives, and applied data analytics in mining, geochemistry, and education. Dr. Keith is an Associate Editor of \emph{Intelligent Data Analysis}. He was a recipient of the Fulbright Faculty Development Scholarship (2019-2021), the Becas Chile Doctoral Studies Scholarship (2019-2023), and was Valedictorian of the 2016 and 2017 graduating classes at UCN.\end{IEEEbiography}

\EOD

\end{document}